\documentclass[preprint,aps,12pt,epsfig]{revtex4} 
\usepackage{graphicx}
\begin{document}

% \noindent
\begin{center}
{\Large \bf Connectivity strategies to enhance the capacity of weight-bearing 
networks}
\end{center}
\begin{center}
T. M. Janaki $^1$ and Neelima Gupte$^2$\\
Department of Physics, Indian Institute of Technology Madras,\\
Chennai-600 036, INDIA.
\end{center}

\begin{center}
{\bf \large Abstract}

\end{center}
 
%\begin{abstract}

The connectivity properties of a weight-bearing
network  are exploited to enhance it's  capacity.
We study a
2-d  network of sites where the
weight-bearing capacity of a given site depends on the capacities of the sites  connected to it in the layers above. 
The  network consists of 
clusters viz. a set of sites connected with each other with the largest such collection of sites being denoted as the maximal cluster.
New connections are made between sites in successive layers 
using two distinct strategies.
The key element of our strategies consists of adding 
 as many disjoint clusters as possible 
to the sites on the trunk $T$ of the maximal cluster. 
 In the first strategy the reconnections start from 
the last layer upwards and stop when
    no new sites are added. In the second case, the reconnections 
start from the top layer and go all the way down to the last layer.
The new networks can bear much higher weights than the original
 networks and
have much lower failure rates.
The first strategy
  leads to a greater enhancement of stability whereas the second  leads to
 a greater enhancement of capacity compared to the original networks.
The original network used here is a typical example of the branching hierarchical class.
However the application of  strategies similar to ours can yield useful results in other types of networks as well.

%\end{abstract}


\hfill

\vskip0.25cm
 e-mail:$^1$janaki@chaos.iitm.ernet.in, $^2$gupte@chaos.iitm.ernet.in  
\newpage

\section{Introduction}

  The study of networks has 
  important  applications in varied branches of science and technology and has therefore 
recently emerged as a  widely
 researched area \cite{Strogatz,Barabasi}. 
Networks of practical importance such as  power
 grids, the internet, 
 traffic networks, cellular and metabolic networks, neural and telephone
networks,
have been extensively studied. Many of these networks have been identified to be scale-free networks \cite{Barabasi,Kim}.
The importance of small-world networks which mediate between regular and random networks has also been recognised\cite{Strogatz}.
Studies of networks of dynamically evolving elements have been extensively used for  pattern formation studies  \cite{Kaneko}, and random graph networks have been used to study granular media \cite{Mehta}. The structure and connectivity properties of such networks, as well as the weight-bearing and traffic handling capacities of their nodes have important consequences for their performance and efficiency. In the case of many networks, 
it is relatively 
easy to change the connectivity properties of the structure, e.g. in the World Wide Web, where pages and links are created and destroyed every
 second, and neural synaptic connections are created and destroyed due to learning and aging processes. It is therefore interesting to see whether the connectivity properties of networks can be exploited to enhance their capacities and
thereby their performance and efficiency. 
In this paper, we use connectivity properties to enhance the weight-bearing capacity  of a network of weight-bearing sites.

\section{The Network}
The network is a 
$2-d$ lattice where a site can be connected to one, both
or none of its neighbours in the layer above, while it has to be connected to
exactly one of its neighbours in the layer below. 
Since each site has two neighbours in the layer below, one of the two
neighbours is randomly chosen for the connection. Thus for a given site
in the layer $D$, the choice of connection between it's left and right neighbours in the $D+1^{th}$ layer corresponds to two distinct realisations of the network.
Also, a site has the capacity to bear 
unit weight if it is not connected to any site in the layer above, and
can  bear weight $W+1$ if it is connected to sites whose capacities add up
to $W$, in the layer above. Therefore, the capacity  $W(i^D)$ of the $i^{th}$
 site in the $D^{th}$ layer is given by:
$$ W(i^D)=l(i_l^{D-1},i^D)~W(i_l^{D-1})+l(i_r^{D-1},i^D)~W(i_r^{D-1}), $$
where $i_l^{D-1}$ and $i_r^{D-1}$ are the left and the right neighbours of $i$ in
$D-1^{th}$ layer. The quantity $l(i_l^{D-1},i^D)$ takes the value $0$ if there is no connection between
$i_l^{D-1}$ and $i^D$ and $1$ if a connection exists. 
We show one realisation of a  network of $64$ sites arranged in eight layers in Fig. 1.
The connections are indicated by lines in the figure. The
weight-bearing capacity of each site is indicated by the number in brackets below the site.
This is the $q(0,1)$
case of the Coppersmith model of granular media and is also a model for river networks
 \cite{coppersmith,river}.
The injection
and aggregation rule of river networks,  by which the  flow at a site
is the sum of injections over all sites upstream of it 
 plus it's own injection,
gives rise to similar
structures. In this context, the capacity could be the reservoir capacity at each site, with each site being considered capable of holding all the water
that comes into it from up-stream as well as it's own injection.

The  network consists of 
clusters viz. a set of sites connected with each other with the largest such collection of sites being denoted as the maximal cluster.
Typical clusters 
$C_1,C_2,C_3,C_4$ are seen in  the realisation in Fig. 1 
with $C_2$ being the maximal cluster.
We look for connectivity strategies which  enhance the weight-bearing 
capacity of the total network. Our strategies involve connecting as many sites as possible from
 various
disjoint clusters to the
 trunk of the maximal cluster.
 We then compare the maximal weight-bearing capacity, the manner of transmission of a weight placed on an arbitrary site in 
the
first layer of the lattice, and the  failure rate of transmissions of the original and the  modified lattices where the failure rate of transmissions is defined as the fraction of transmissions which reach a site which can neither take the weight transmitted to it, nor transmit it to neighbouring sites.
The technique used by us is extremely successful in enhancing the capacity of the network, and also results in a substantial reduction in the failure rate of transmissions. Our techniques are general and could yield similar results in other network models as well.

The weight transmission in the network takes place along
  the connections between  sites which serve as paths. When a site
in the first layer of the network receives a weight ${\cal W}$, 
it retains an amount equal to its
 capacity $W$ and transmits the rest, i.e. ${\cal{W}}-W$, to the site it is 
connected to, in the layer below. Hence the weight transmission is in 
the downward direction and the sites involved in this process
with their connections constitute the path of transmission. Let $P$ 
be one such path and $P_{D}$ be the site on $P$ in the $D^{th}$ layer..
Then, the excess weight at a site $P_D$ in the $D^{th}$ layer is given by:
$$W^{ex}(P_D)= {\cal{W}}-\sum_{K=1}^{D}~W(P_K).$$ 
If $W^{ex}(P_D) \le 0$, then the transmission ends at the $D^{th}$ layer
of the path $P$ and is considered to be successful. On the other hand,
if $W^{ex}(P_D) > 0$, the weight is transferred to $P_{D+1}$. 
Finally, if there is still excess weight left at the $M^{th}$ layer, it
is then transmitted to the corresponding site in the first layer and the
second cycle of downward transmission begins as described above.
 This process of weight transmission, defined as an avalanche, continues
 till either there is no excess weight left, which is defined as a successful transmission,
or  the receiving site is not able to transmit the excess to
the site in the layer below.  
 This occurs when it is
 connected to a site that has already received its share of the
weight (i.e. saturated it's capacity) in the first cycle of transmission,
 thus making further transmissions impossible. Such a transmission 
is said to have failed.
 The time
taken for an avalanche is defined as the number of layers traversed by
the weight in the network.

To test for avalanches in our study, the weight to be placed on any site in the first layer
is chosen to be the sum of the weight-bearing capacities of sites along a path which we call the trunk of the maximal cluster (see beaded lines in $C_2$ in Fig. 1). Let the site on the $D^{th}$ layer of the trunk be denoted by
$T_D$ ($D=1, \ldots, M$), where the $M^{th}$ layer is the last layer of the lattice.
To obtain the sites of the trunk as follows: We choose $T_M$ as the site with
 the maximum
 capacity in the $M^{th}$ layer ($T_8$ is the $4^{th}$ site in the $8^{th}$ layer in Fig. 1).
 Clearly, $T_M$  belongs to the maximal cluster. Of its two neighbours in the $M-1^{th}$ layer,
 we chose the one which is connected to $T_M$ and denote it by $T_{M-1}$. If both the sites are connected,
 then the one that lies on a path running through the entire height of the network
is chosen.
If both lie on such paths, then $T_{M-1}$ is chosen to be the site
 with maximum 
capacity ($T_7$ is the $5^{th}$ site in layer $7$ in Fig 1). 
We repeat this process 
till we reach the first layer and obtain all the $T_D,~D=1, \ldots, M$.
Let $W_T = \sum_{D=1}^{M}~W(T_D)$, the sum of the capacities of the 
sites of the trunk ($W_T=64$ for Fig. 1).
Clearly, a weight equal to $W_T$ can be transmitted successfully in the network,
 if it is placed on $T_1$ but may result in  transmission failure if 
placed at other sites in the $1^{st}$ layer. 
If  a weight ${\cal W}=W_T$ is placed on an arbitrary site in the first layer of the
 network, the number of failed transmissions for networks 
of sizes $N=50 \times 50$, $75 \times 75$, $100 \times 100$ and $150 \times
150$ is given in the first column of Table I. We see that almost 50 \% of the transmissions result
in failure.

\section{Capacity enhancement strategies}

We now look for ways to increase the weight-bearing capacities of the sites in the
network. Our method consists of re-connecting a site on a given layer to a site of our
choice in the layer below. The sites can been chosen in different ways. The reconnections are restricted to at most
one per layer with the total number of connections kept constant. As
any weight-propagation or avalanche tends to take place along a connected path, it is beneficial to make the new connections
to the sites which lie on a path in a cluster, rather than to arbitrary sites.

In the first strategy (Strategy I), we connect as many disjoint clusters as possible 
to the sites on the trunk $T$ of the maximal cluster 
so that the maximum number of sites are included in the cluster.
Hence, we choose from the penultimate  layer
 a site $I_{M-1}$,
such that it does not belong to the maximal cluster and whose degree is $3$
(i.e. it is connected to three sites)
 (the $2^{nd}$ site in the $7^{th}$ layer in Fig. 2). We
snap off its existing connection to the site in the $M^{th}$ layer to
 reconnect it to the site $T_M$ on the trunk. 
If there is more than one such site, we choose $I_{M-1}$ as the site, whose
shorter branch has the maximum  capacity in the layer above.
On the other hand, if there are no such sites, then 
the site with maximum capacity and
which does not belong to the maximal cluster is chosen for the reconnection. 
Thus  $W(T_M)$ gets enhanced by an
amount equal to $W(I_{M-1})$. Similarly, we chose $I_{D}$
in the $D^{th}$ layer and connect it to $T_{D+1}$. Therefore the capacity
enhancement is given by,   
$W(T_{D})=W(T_{D})+\sum_{D^\prime} W(I_{D^\prime})$,${D^\prime}=D,D-1,\ldots L$
i.e.  the sum runs over all the sites chosen by strategy I in the previous 
layers with  the process coming to an end at the layer $L$ when  we fail to add any new site 
to the maximal cluster. 

As we  connect as many disjoint clusters as possible to the maximal 
cluster, 
any site on the first layer gets connected to the trunk
at some layer. Therefore, these reconnections achieve the dual objective of enhancing
the weight-bearing capacities of the sites on the trunk as well as making the trunk 
accessible from any site on the first layer. Hence, any weight placed on the first layer
reaches the trunk at some layer, which with its enhanced capacity has a greater probability of supporting the weight successfully than other paths in the network.
The enhancements obtained in the sum of the  capacities
of sites on the trunk of the maximal cluster i.e. the new capacity $W_T^{enh}$
in  modified networks of different sizes averaged over $1000$ realisations
are listed in  Table II relative to the capacity of the original trunk. 
It is clear that we obtain a huge enhancement in the weight-bearing capacities
of the networks which increases with the size of the network, as expected.
We plot the dependence of the percentage increase in capacity against
the number of layers in Fig 3. It is clear from the plot that
the increase in capacity $\Delta W_T^{enh} \approx  ln \phantom{a}M$.
We also examine the stability of the new networks to weight-transmission.
The  weight $W_T^{enh}$ is placed on a randomly chosen site in the
first layer and allowed to propagate. 
The failure rates of the modified networks of different sizes
are listed in Table I.
We find a substantial reduction in the number
of failed transmissions in the networks modified by Strategy I compared to the original network. 
The  failure
rate drops from almost 50 \% in the original networks to around 10 \% in the
modified networks (see Table I), while the weight-bearing capacity increases substantially. 
This reduction is the result of 
the manner in which the reconnections are introduced.
As we stop making the reconnections when we fail to add any new site to the maximal cluster, the reconnections 
are restricted to the lower layers only, leaving the upper layers 
undisturbed. Therefore, the weight has more layers to traverse and distribute
itself among the sites before being forced onto
the trunk in each cycle of its propagation. This increases the chances of its
successful transmission enormously. This strategy is therefore successful in
increasing the stability as well as the weight-bearing capacity of the networks.

To enhance the weight-bearing capacity of the network further,
a natural way  would be to start the reconnections from the first layer onwards 
so that the  capacities of the site on the trunk in the layer below and its subsequent
sites get enhanced in every layer. This is  Strategy II.
This strategy achieves maximum enhancement in the  capacities of the sites
on the trunk as the reconnections are
introduced from the
first layer itself (See Fig. 4). 
 Here, at every layer $D$, a site which does not belong
to the maximum cluster and which has the maximum
capacity is chosen for the reconnection and denoted by ${II}_D$.
 Therefore, after the reconnection, the new capacity of the sites
$T_D$ of the trunk is given by:
$W(T_{D^{\prime}})= W(T_{D^{\prime}})+W({II}_D)$,where $D^{\prime}=
2,\ldots, M$. 
 Therefore, $W_T=\sum_{D=1}^M W(T_D)$, 
 gets enhanced,
 resulting in the maximum
enhancement of the capacity of the trunk that can be achieved 
with the restriction of one reconnection per layer.
Note that each reconnection changes the capacities of the layer 
below and the new capacities have to be taken into account before a 
new site is chosen for the reconnection in the next layer. While a similar change takes place in the case of Strategy I as well, the site chosen for the reconnection does not change as the reconnections start from the bottom layer upwards except for the fact that no reconnections are made to sites which belong to clusters 
which have been reconnected to the maximal cluster at lower layers.
The increase in the sum of the weight-bearing capacities of the trunk $T$ of
 the maximal cluster in the modified networks of different sizes
relative to that of the
 unmodified networks, averaged over $1000$ realisations is listed in Table I.
 This strategy gives the maximum enhancement that can be achieved with
the restriction of one reconnection per layer. As before, $\Delta W_T^{enh}$ the percentage 
increase in capacity depends on the size of the network, however, it can be seen from Fig. 5  that the log-log plot of $\Delta W_T^{enh}$ versus $M$ can be fitted to a straight line, so that $\Delta W_T^{enh} \approx M^\alpha$, where $\alpha \approx 0.66$.
 
We  study the effect of this strategy on weight transmission when a weight 
${\cal W}= W_T^{enh}$ is placed on a randomly chosen site in the first layer. 
Similar to Strategy I, any site on the first layer gets connected to 
the trunk at
some layer. This is because
 if a site $i$ on the first layer
remains unconnected to the trunk of the
maximal cluster for some time, the capacity of a site on the path
originating from $i$
 becomes 
maximum in some layer and gets reconnected to the 
the trunk . The sites of the maximal cluster are
 connected to the trunk at some layer.
Hence a weight placed on any site in the first layer reaches the
trunk of the maximal cluster eventually through some reconnection.  
Unlike  Strategy I, the weight 
 reaches the trunk faster as the reconnections start from the
first layer itself. Hence, the weight does not
 traverse many layers before reaching the trunk in a cycle of its
propagation so that the number of possible new paths in each cycle is reduced. This results in a significant
 decrease in the number of successful transmissions compared to Strategy I,
though there is  a substantial increase over the stability of 
the original (see Table I).
On the other hand, this strategy shows an enormous increase in the weight-bearing capacity 
of the network, while Strategy I leads to a lower failure rate \cite{fn2}.

The probability distribution of avalanche durations for strategy 
is  shown in Figs 6, 7 and 8 for an ensemble of $2100$ successful weight transmissions
 for a weight $W_T^{enh}$  placed
at a random site in the first layer for the original lattice (Fig. 6), for strategy I distributions (Fig. 7) and for Strategy II distributions (Fig. 8). 
No  avalanches where $t/M< 1$ are seen as there are no paths
with capacities greater than the $W_T^{enh}$, in any of the three cases and
the original networks have avalanches which can cycle thrice through the network.
However, in the case of strategies I and II,
no avalanches of length $t/M >2$ are seen
as
every site in the first layer gets eventually connected to the trunk and the weight in its second
cycle of propagation  reaches a site on the trunk where it can settle down
or where transmission fails.
 The distributions for networks of different sizes collapse on one another when they are scaled by their
respective number of layers for all three cases as can be seen in Figs. 6,7  and 8\cite{FN}. 

\section{Discussion}

A practical example of a situation where our ideas could be applicable is that
of grid computing on computers connected in a branching hierarchical manner
with connections to a central backbone. A task dumped on any arbitrary computer
at the first level finds it's way to the central backbone, which contains computers of high computational capacity. The task is distributed parallely along the
path with each computer on the path processing the task according to it's available capacity. The original network is capable of handling tasks of a certain magnitude without jamming the network. Given a task of a higher magnitude our strategies permit us to rerout our task along the network with suitably enhanced capacities at a few nodes so that the network becomes capable of handling the given
task without jamming. We also note that since most networks have finite extent,
once a task reaches the end of the network without finding a node with adequate capacity to handle it, it is useful to find  a simple re-routing which can find a new path where it might settle down rather than dumping the task. Recycling
constitutes such a simple re-routing.  
 We  note that once recycling is done most weights or tasks settle down in less than three cycles or find a node which fails leading to the collapse of the network. Hence recycling is a
worthwhile strategy to pursue.

Thus we have identified a set of strategies which enhance the weight-bearing capacity of a branching hierarchical network of weight-bearing sites.
The  capacity of the sites in a cluster increases according to the number of sites that they are connected to  above them.  The maximal cluster, which connects the largest number of sites  contains the  sites and paths which possess high weight-bearing capacities. Therefore, strategies which  connect
as many sites as possible from
disjoint clusters to the
 trunk of the maximal cluster  successfully enhance the weight-bearing properties of the cluster. Since the trunk of the maximal cluster can take large weights, the addition of extra connections to this result in the transferrence of the weight to the path which can bear it most successfully,  resulting in a drastic reduction of the failure rate of the structure. The stability of the structure
is thus greatly enhanced. The enhancement in the weight-bearing properties
is largest for the strategy which connects to the trunk in the higher layers,
on the other hand, the stability is enhanced by allowing avalanches to propagate for a few layers before connecting to the main trunk. However, both strategies result in enormous  enhancements of the weight-bearing capacity and the
stability over the original network. We note that the improved properties of the new network are achieved by adding connections to the strongest sites in the network, viz. the sites which belong to the trunk. We expect that this feature will carry over to any network which contains sites which are of significantly higher capacity than the average site in the network. 
It does not appear that the specific rule  used by us for  the addition of weight-capacities is essential for total capacity enhancement.
We also note that sites which have more connections than others contribute more to capacity enhancement.
This is a feature which has been noted in other contexts e.g. it has been observed that search algorithms which exploit high connectivity nodes function more effectively\cite{Huberman}. Our strategies have been tailored to enhance the weight-bearing capacity of our network. However they could be exploited to enhance the
information or traffic carrying capacities of communication networks or to improve navigability in networks\cite{Kleinberg}.
We hope to examine some of these issues in future work.
\\

\noindent {\large  \bf Acknowledgements:} We thank A. Mehta, O. Narayan and K. Jain for extensive discussions and CSIR (India) for partial financial support.

%\vskip0.2in
%\hline
%\vskip0.2in

\newpage
\begin{center}
\vskip 0.1in
\noindent {\bf Table I:} The failure rates of weight transmission in 
the original and
 modified networks.

\vskip0.2in

\begin{tabular}{|l|c|c|c|}
\hline
Network size & Original & Strategy I & Strategy II \\ \hline
$N=50 \times 50$ & $49.2 \%$ & $8.3 \%$ & $26.2 \%$ \\ \hline
$N=75 \times 75$ & $48.5 \%$ & $9.8 \%$ & $27.2 \%$ \\ \hline
$N=100 \times 100$ & $49.7 \%$ & $11.9 \%$ & $23 \%$ \\ \hline
$N=150 \times 150$ & $51.9 \%$ & $9.1 \%$ & $25.5 \%$ \\ \hline
\end{tabular}
\end{center}
\vskip1cm

\noindent {\bf Table II:} The enhancement in the weight-bearing capacities of 
trunks of the
modified networks
with respect to their original networks.

\vskip0.2in

\begin{center}
\vskip 0.1in
\begin{tabular}{|l|c|c|c|}
\hline
Network size & Original & Strategy I & Strategy II \\ \hline
$N=50 \times 50$ & $100 \%$ & $184.1 \%$ & $548.65 \%$ \\ \hline
$N=75 \times 75$ & $100 \%$ & $209.08 \%$ & $721.52 \%$ \\ \hline
$N=100 \times 100$ & $100 \%$ & $232.65 \%$ & $892.6 \%$ \\ \hline
$N=150 \times 150$ & $100 \%$ & $257.025 \%$ & $1129.40 \%$ \\ \hline
\end{tabular}
\vskip 0.5in
\end{center}
\newpage

\begin{figure}
\includegraphics[width=14truecm,height=8cm]{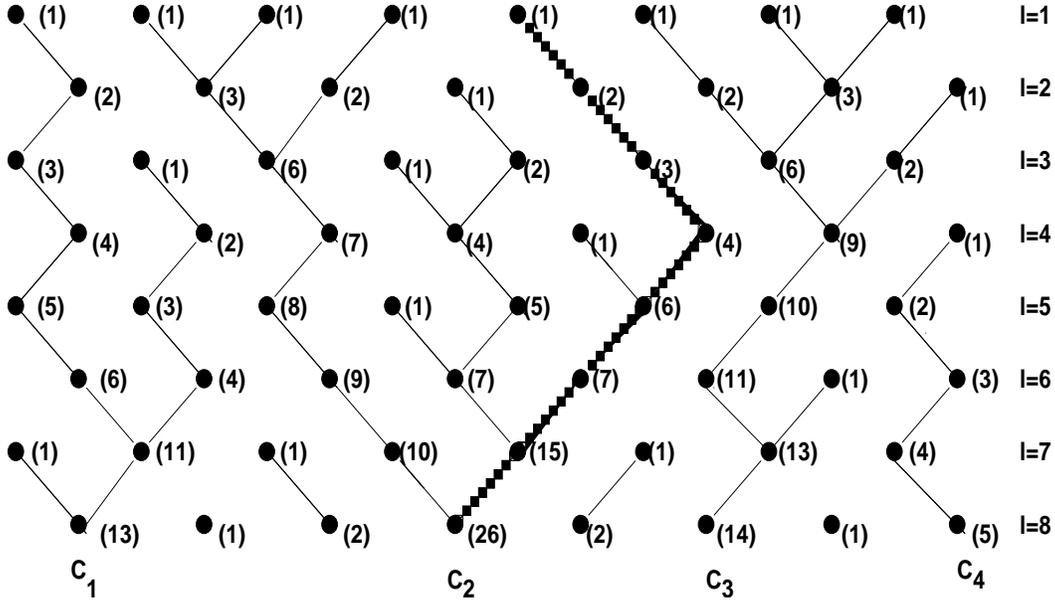}
\caption{\label{fig:epsart8}A network of  $M=8$ layers with $8$
sites per layer.
$C_2$ is the maximal cluster. The beaded line is trunk of the maximal cluster. The weight-bearing
capacity of the trunk is $W_T=64$.}
\end{figure}

\begin{figure}
\includegraphics[width=14truecm,height=8cm]{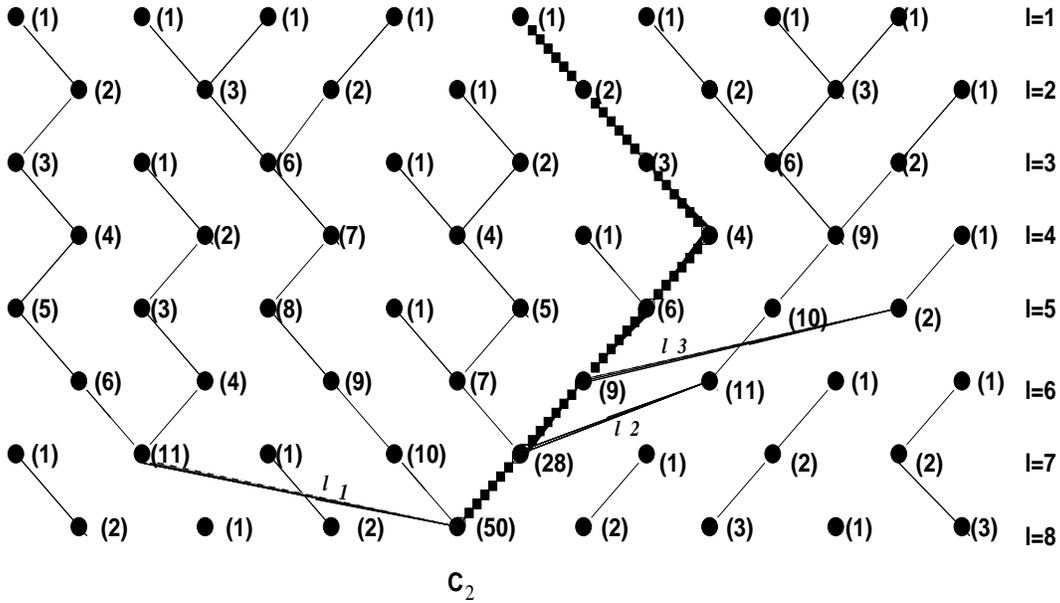}
\caption{\label{fig:epsart1}
A Strategy I network. The links ${\it l1}$, 
${\it l2}$ and ${\it l3}$ are  reconnections to the last three sites on the
trunk of the maximal cluster $C_2$. The new
capacity of the trunk $W_T^{enh}=103$.}
\end{figure}

%\newpage
\begin{figure}
\includegraphics[width=11truecm,angle=270]{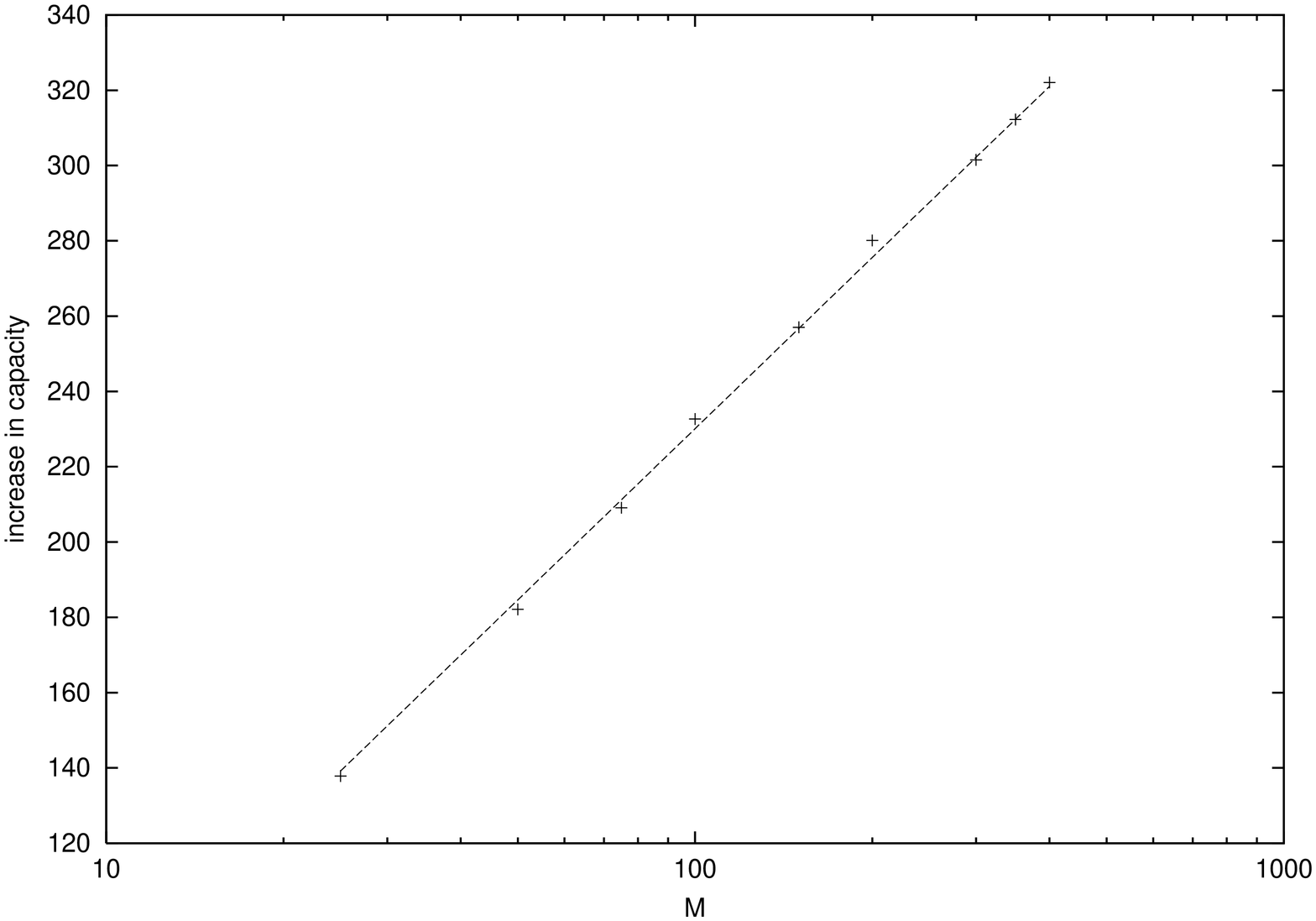}
\caption{\label{fig:epsart2} 
Plot of increase in capacity $\Delta W_T^{enh}$ (dimensionless units) versus $M$, the number of lattice sites for Strategy I networks
(logscale on x-axis).} 
\end{figure}

\begin{figure}
\includegraphics[width=14truecm,height=8cm]{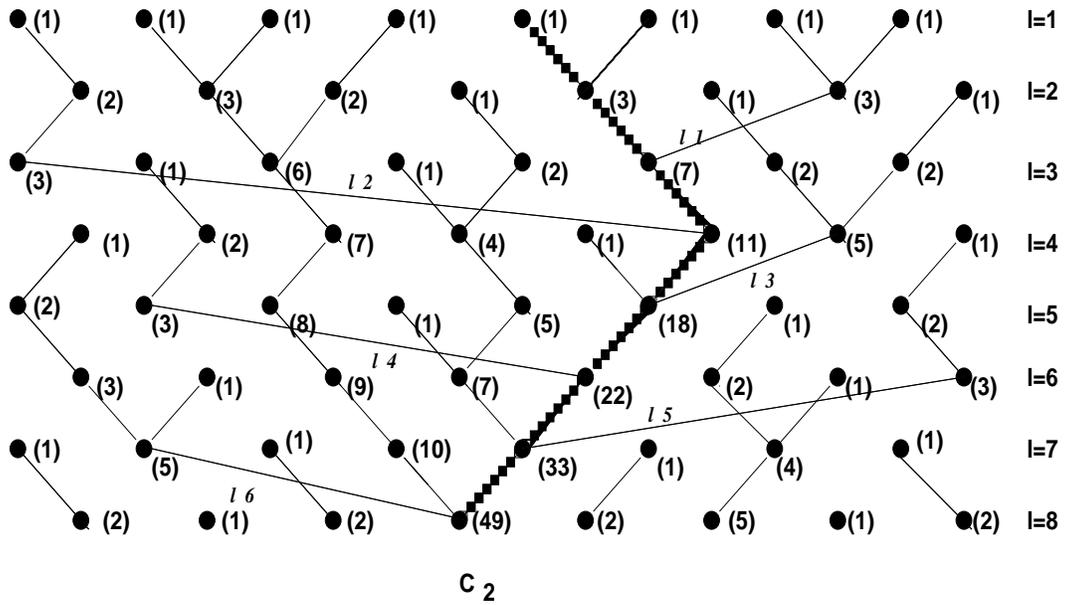}

\caption{\label{fig:epsart3}
A Strategy II network. The links
${\it l1}$, ${\it l2}$, ${\it l3}$, ${\it l4}$, ${\it l5}$ and ${\it l6}$
are  reconnections to the trunk of the maximal cluster $C_2$
from the second layer
onwards. $W_T^{enh}=144$.}
\end{figure}

\begin{figure}
\includegraphics[width=11truecm,angle=270]{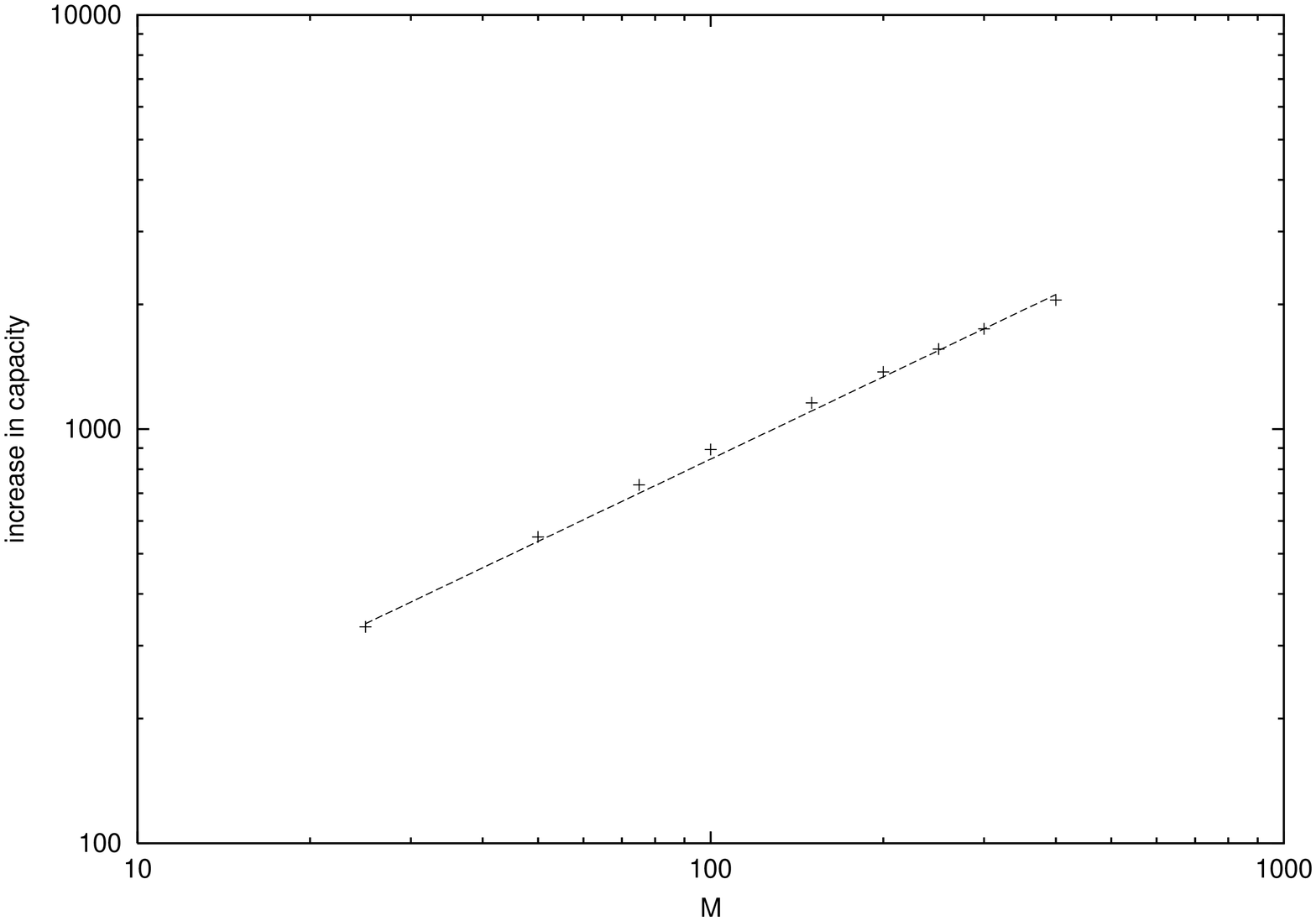}

\caption{\label{fig:epsart4}
Plot of increase in capacity $\Delta W_T^{enh}$ (dimensionless units) versus $M$, the number of lattice sites,  for Strategy II networks
 (logscale on x and y axes).}
\end{figure}

\begin{figure}
\includegraphics[width=11truecm,angle=270]{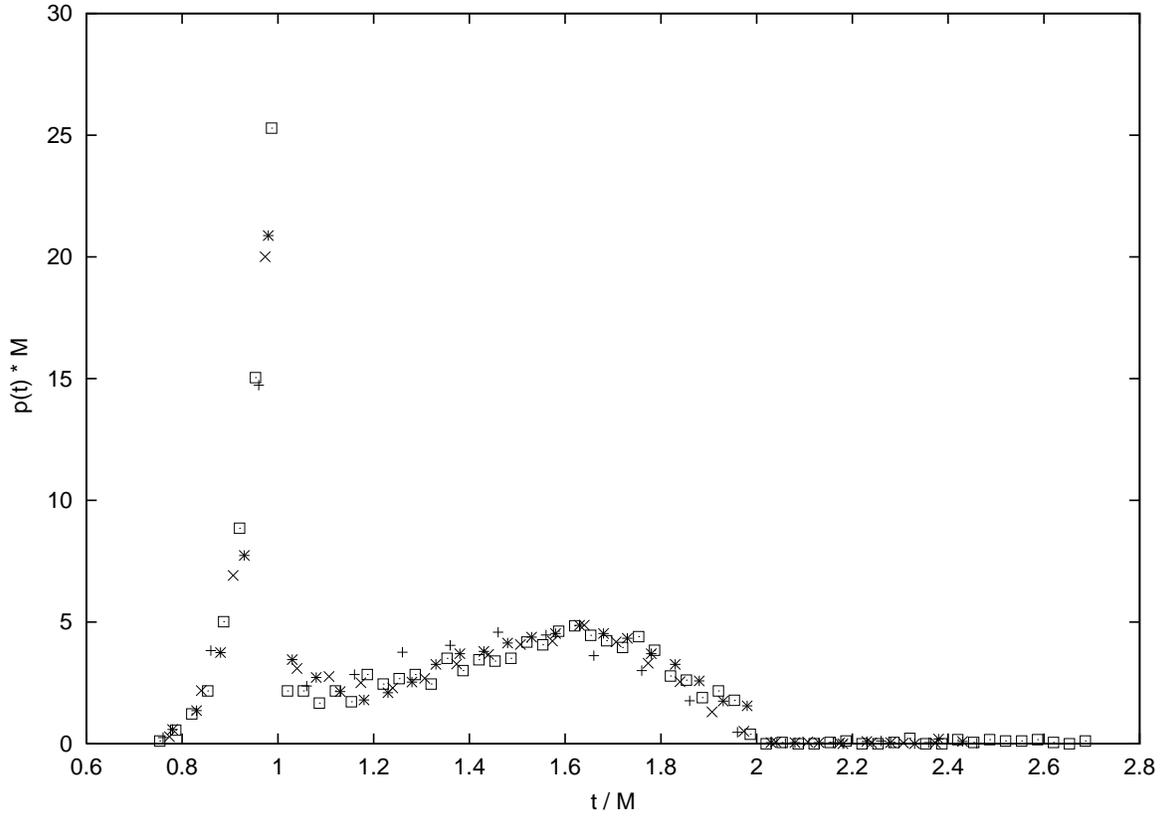}
\caption{\label{fig:epsart5}
The probability
distributions of avalanche times $t$
for $2100$ realisations of the original networks. Data for lattice sizes $N=50 \times 50$ is indicated by plus signs,
$N= 100 \times 100$ by asterisks and $N=150 \times 150$
by squares.  The y-axis is scaled by $M$, the number of layers. 
The x-axis is $t/M$ which is the number of times a test weight cycles through the lattice layers.
The same convention is followed
for different lattice sizes in Fig. 7 and Fig. 8.}
\end{figure}

\begin{figure}
\includegraphics[width=11truecm,angle=270]{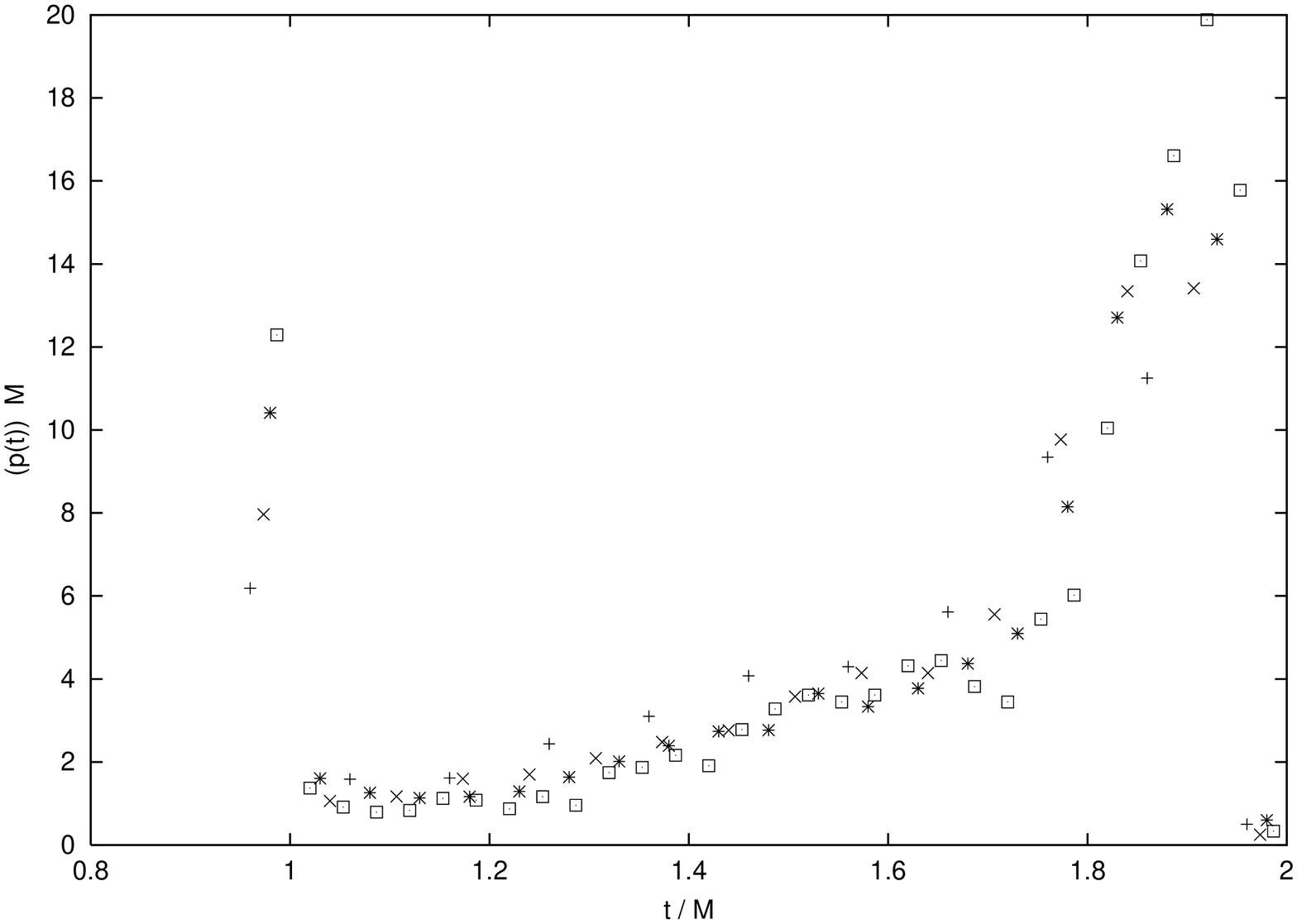}
\caption{\label{fig:epsart6}
The probability 
distributions of avalanche times $t$
for $2100$ realisations of Strategy I networks.} 
\end{figure}

\begin{figure}
\includegraphics[width=11truecm,angle=270]{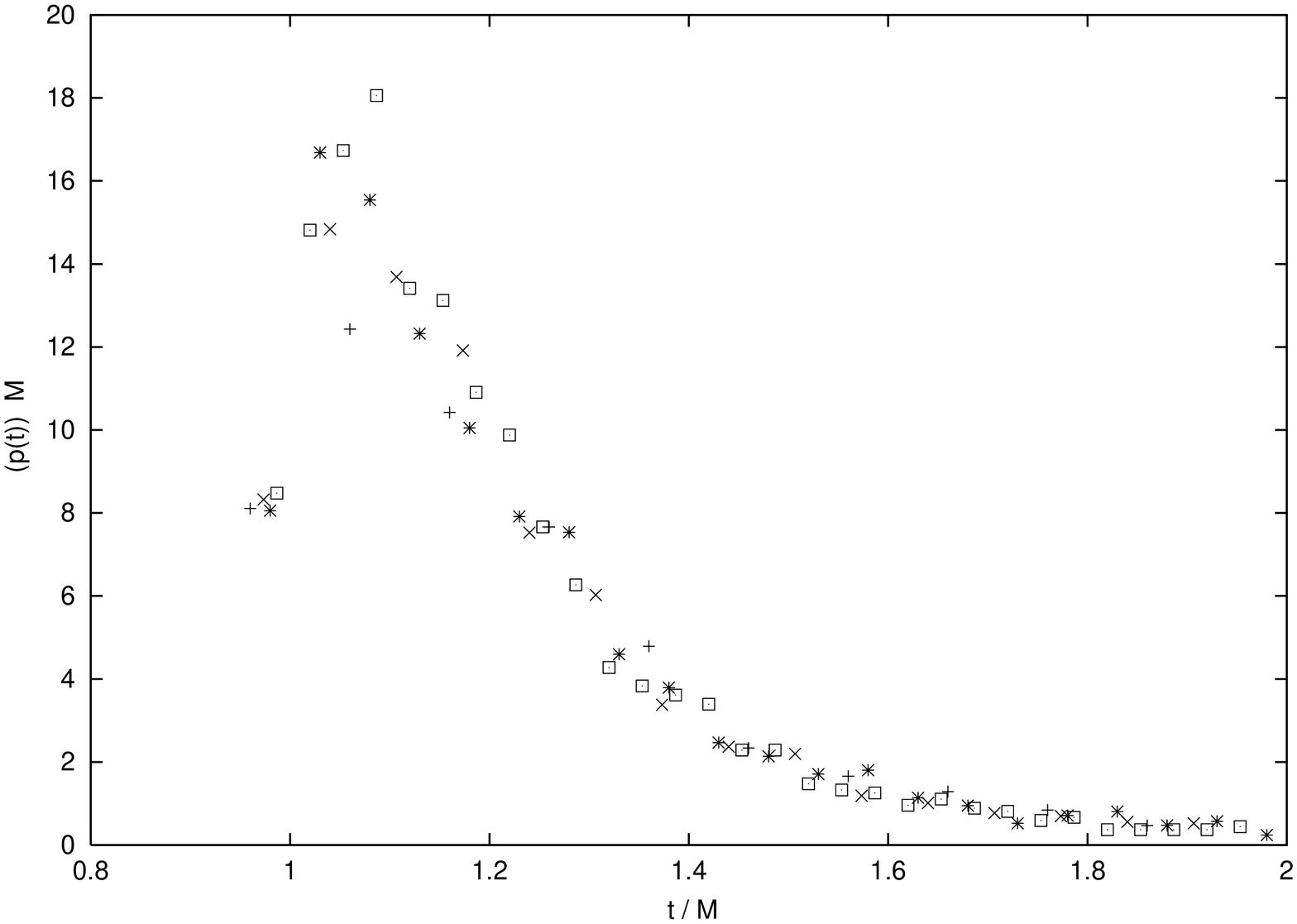}
\caption{\label{fig:epsart7}
The probability 
distributions of avalanche times $t$
for $2100$ realisations of Strategy II networks.} 
\end{figure}

\begin{thebibliography}{99}
\bibitem{FN}

The apparent increase in the height of the peak in the distributions is due to the distribution being scaled by the system size.
The actual fluctuation is within one standard deviation for each distribution.
\bibitem{fn2}

If weights which are higher than the trunk capacity are placed on arbitrary 
sites on the front row the networks show higher failure rates than if they are tested with weights equal to the trunk capacity. Thus the original networks 
fail $50 \%$ of the time when tested with weights $W_T$ and $60 \%$ of the time when tested with weights which exceed $W_T$ by $10 \%$. Strategy I networks 
fail $10 \% $ of the time when tested with weights $W_T^{enh}=1.8 W_T$ and $30
 \%$ of the time when tested with 
weights which exceed $W_T^{enh}$ by $10 \%$ (and thus $W_T$ by $98 \%$), whereas strategy II networks fail about $25 \%$ of the time when tested with weights 
$W_T^{enh}=5.5 W_T$ and $60 \%$ of the time when tested with weights which 
are $10 \%$ more than $W_T^{enh}$ (and thus $W_T$ by $500  \%$. Thus the features of increased capacity and stability are demonstrated again by the new networks with strategy I networks      having  greater stability and strategy II networks having        greater load carrying capacity.
\end{thebibliography}
\end{document}